\documentclass[aps,pra,showpacs,amssymb,superscriptaddress,nofootinbib,twocolumn]{revtex4-1}

\usepackage{subfigure}
\usepackage{dsfont}
\usepackage{color}
\usepackage{dcolumn}
\usepackage{bm}
\usepackage{amsmath}
\usepackage{braket}
\usepackage{amsmath,amsfonts,amsthm,amssymb}
\usepackage{graphicx}
\usepackage[bookmarks=true, colorlinks]{hyperref}

\newcommand\sh{$^2$S$_{1/2}\,$}
\newcommand\ph{$^2$P$_{1/2}\,$}
\newcommand{\qdown}{$^2$S$_{1/2}\,(F=0)\,$}
\newcommand{\qup}{$^2$S$_{1/2}\,(F=1, m_F=+1)\,$}

\newcommand{\ybplus}{$^{171}$Yb$^+\,$}

\newcommand{\E}{\ensuremath{{\mkern1mu\mathrm{e}\mkern1mu}}}

\newcommand{\fig}[1]{Fig.~\ref{#1}}
\newcommand{\eq}[1]{Eq.~(\ref{#1})}
\newcommand{\Sec}[1]{section~\ref{#1}}
\newcommand{\mean}[1]{\left\langle\,#1\,\right\rangle}

\begin{document}

\title{Radio-frequency sideband cooling and sympathetic cooling of trapped ions in a static magnetic field gradient}

\author{Theeraphot Sriarunothai}
\affiliation{Department Physik, Naturwissenschaftlich-Technische Fakult{\"a}t, Universit{\"a}t Siegen, Walter-Flex-Str. 3, 57068 Siegen, Germany}

\author{Gouri Shankar Giri}
\affiliation{Department Physik, Naturwissenschaftlich-Technische Fakult{\"a}t, Universit{\"a}t Siegen, Walter-Flex-Str. 3, 57068 Siegen, Germany}

\author{Sabine W{\"o}lk}
\affiliation{Institute for Theoretical Physics, University of Innsbruck, Technikerstra{\ss}e 21a, 6020 Innsbruck, Austria}
\affiliation{Department Physik, Naturwissenschaftlich-Technische Fakult{\"a}t, Universit{\"a}t Siegen, Walter-Flex-Str. 3, 57068 Siegen, Germany}

\author{Christof Wunderlich}
\email{wunderlich@physik.uni-siegen.de}
\affiliation{Department Physik, Naturwissenschaftlich-Technische Fakult{\"a}t, Universit{\"a}t Siegen, Walter-Flex-Str. 3, 57068 Siegen, Germany}

\begin{abstract}
We report a detailed investigation on near-ground state cooling of one and two trapped atomic ions. We introduce a simple sideband cooling method for confined atoms and ions, using RF radiation applied to bare ionic states in a static magnetic field gradient, and demonstrate its application to ions confined at secular trap frequencies, $\omega_z \approx 2\pi\times 117 $kHz. For a single \ybplus ion, the sideband cooling cycle reduces the average phonon number, $\mean{n}$ from the Doppler limit to $\mean{n} =$ 0.30(12).
This is in  agreement with the theoretically estimated lowest achievable phonon number in this experiment. We extend this method of RF sideband cooling to a system of two \ybplus ions, resulting in a phonon number of $\mean{n} =$ 1.1(7) in the center-of-mass mode. Furthermore, we demonstrate the first realisation of sympathetic RF sideband cooling of an ion crystal consisting of two individually addressable identical isotopes of the same species.      

\end{abstract}
\maketitle


\section{Introduction}
\label{sec:introduction}

Cooling trapped atomic ions and neutral atoms close to their vibrational ground state is often a prerequisite for experiments in quantum optics and quantum information science. Cooling beyond the Doppler limit, eventually leading to the vibrational ground state, using sideband cooling was experimentally demonstrated initially using an electric  quadrupole transition in a single $^{198}$Hg$^+$ ion \citep{diedrich1989} and later using a Raman transition in a single $^9$Be$^+$ ion \citep{monroe1995}. Since then several experiments using the resolved sideband cooling technique were carried out and, for example,  vibrational ground state occupation probabilities of 99.9\% \citep{roos1999} and 99\% \citep{thompson2013} have been achieved, for ions at an axial trap frequency of $2\pi\times 4.51$ MHz and for neutral atoms at a radial trap frequency of $2\pi \times 100$ kHz, respectively.

For an efficient realisation of (near-)ground state cooling in typical traps, atomic transitions in the optical regime have been a natural choice, since they provide a sufficiently large  Lamb-Dicke parameter (LDP), a measure for the strength of coupling between internal and vibrational states of a trapped ion or atom \citep{eschner2003,segal2014}. Recent years have witnessed the emergence of long-wavelength radiation in the radio frequency (RF) regime as a promising alternative to generate the required coupling between internal and vibrational states. While RF radiation is relatively easy to generate and to control as compared to optical  radiation, the application to trapped ions and atoms is characterised by a too small LDP for most practical purposes at usual trap frequencies. This problem can be circumvented with the help of an additional static or oscillating magnetic field gradient \citep{mintert2001,ospelkaus2008,welzel2011,woelk2017,weidt2016, harty2016,piltz2016,yuji2017,wahnschaffe2017}. Near-ground state cooling employing RF radiation has been previously demonstrated using an oscillating magnetic field gradient \cite{ospelkaus2011} or a static magnetic field gradient \citep{scharfenberger2012,weidt2015}. In the former implementation, sideband cooling is performed on a radial vibrational mode (2$\pi \times$ 4-10 MHz), while the latter implementation makes use of an axial mode (2$\pi \times$ 427 kHz) using atomic states dressed by RF fields.  

Here we report a detailed investigation of sideband cooling trapped \ybplus ions using RF radiation applied to bare ionic states in  a static magnetic field gradient. We cool individual ions and two-ion Coulomb crystals to near their motional ground state. The ions are confined in a harmonic trapping potential characterised by a relatively low axial secular frequency, $\omega_z =2\pi\times$ 117 kHz \citep{scharfenberger2012, andres2013}. This technique not only showcases experimental simplicity avoiding the necessity of additional dressing fields, but also promises applications in other physical systems, such as ground state cooling of neutral atoms, where trapping is often achieved at comparable or even lower axial secular frequencies \citep{perrin1998, winoto1999, kaufman2012, reiserer2013}. Since the average achievable phonon number as a function of axial secular frequency $\omega_z$ can be expressed as $\mean{n} \approx T k_B / \hbar \omega_z$, where $T$ is the temperature of an ion and $k_B$ is the Boltzmann constant \citep{stenholm1986}, the presented technique can be expected to work even better for higher axial secular frequencies. In addition, we investigate sympathetic cooling of one \ybplus by a second, identical \ybplus ion made possible by individually addressing  each ion with RF radiation. 

The paper is organised as follows. In \Sec{sec:experimental-setup} we briefly describe the experimental apparatus. In \Sec{sec:mw-sbc} we discuss the principle of RF sideband cooling and report its implementation in one- and two-ion systems. Finally we summarise our results in \Sec{sec:conclusion}. Supplementary mathematical details are collated in the appendices.



\section{Experimental setup}
\label{sec:experimental-setup}

The experiments reported here were carried out with one or two laser cooled \ybplus ions confined in a linear Paul trap \cite{khromova2012,piltz2016}. Qubit states are encoded in the hyperfine manifold of the ground \sh ~level (see \fig{level-scheme}). The \qdown ~state represents state $\ket{0}$ and the \qup ~state represents state $\ket{1}$. Doppler cooling is achieved by optically driving the $^2$S$_{1/2}(F=1) \leftrightarrow ~^2$P$_{1/2}(F=0)$ transition using 10 MHz red-detuned laser light near 370 nm and, simultaneously,  driving the RF resonance $\ket{0} \leftrightarrow \ket{0'}$  near 12.6 GHz, where $\ket{0'}\equiv \   $\sh$(F=1,m_F=0)$.  Both fields are turned on for a duration of 5 ms. Two additional lasers with wavelengths near 935 nm and 638 nm are employed to repump the ion back from metastable states $^2$D$_{3/2}$ and $^2$F$_{7/2}$ (not shown in \fig{level-scheme}) to the Doppler cooling cycle \cite{scharfenberger2012,bell1991,engelke1996}. 

The latter transition is sensitive only in second order to the magnetic field strength, and hence the differential Zeeman shift between two neighbouring ions is small ($2\pi \times$ 8 kHz) compared to the Rabi frequency ($2\pi \times$ 45.9 kHz), thereby enabling all ions in the Coulomb crystal to participate in the Doppler cooling cycle. 

The $\ket{0}$ state is prepared by optical pumping into the $^2$S$_{1/2}(F=0)$ state by driving resonantly the $^2$S$_{1/2}(F=1) \leftrightarrow\  ^2$P$_{1/2}(F=1)$ transition. 
State dependent readout is achieved by detecting resonance fluorescence with the exciting laser light being resonant with the cooling transition \cite{woelk2015}. Scattered light can either be detected with a photomultiplier tube, or with an Electron Multiplying CCD (EMCCD) camera, or both. 

The quantization axis is defined along the trap axis by a bias magnetic field of 0.6 mT. A magnetic gradient of 19 T/m along the same direction is generated by two permanent magnets mounted around the end cap electrodes of the trap. This gradient allows addressing of individual qubits in frequency space, since the qubit resonances are individually Zeeman shifted as they depend on the magnetic field at the site of each ion \cite{segal2007,johanning2009,belmechri2013,piltz2014}. Furthermore, magnetic gradient induced coupling (MAGIC) results in a pairwise spin-spin interaction, making the string of ions a pseudo-molecule with a long-range J-type coupling \cite{wunderlich2002,khromova2012,piltz2016}. The axial secular frequency in the experiments presented here is about $2\pi \times$ 117 kHz. The effective LDP  for a single ion in our experiment is ${\eta_{eff}}$ = 0.0359. The measured Rabi frequency, $\Omega_0$ on the $\ket{0} \leftrightarrow \ket{1}$ RF transition in the experiments reported here is 2$\pi \times$ 39.47(4) kHz.



\section{Radio-frequency sideband cooling}
\label{sec:mw-sbc}

\begin{figure}[]
\centering
\includegraphics[width=0.47\textwidth]{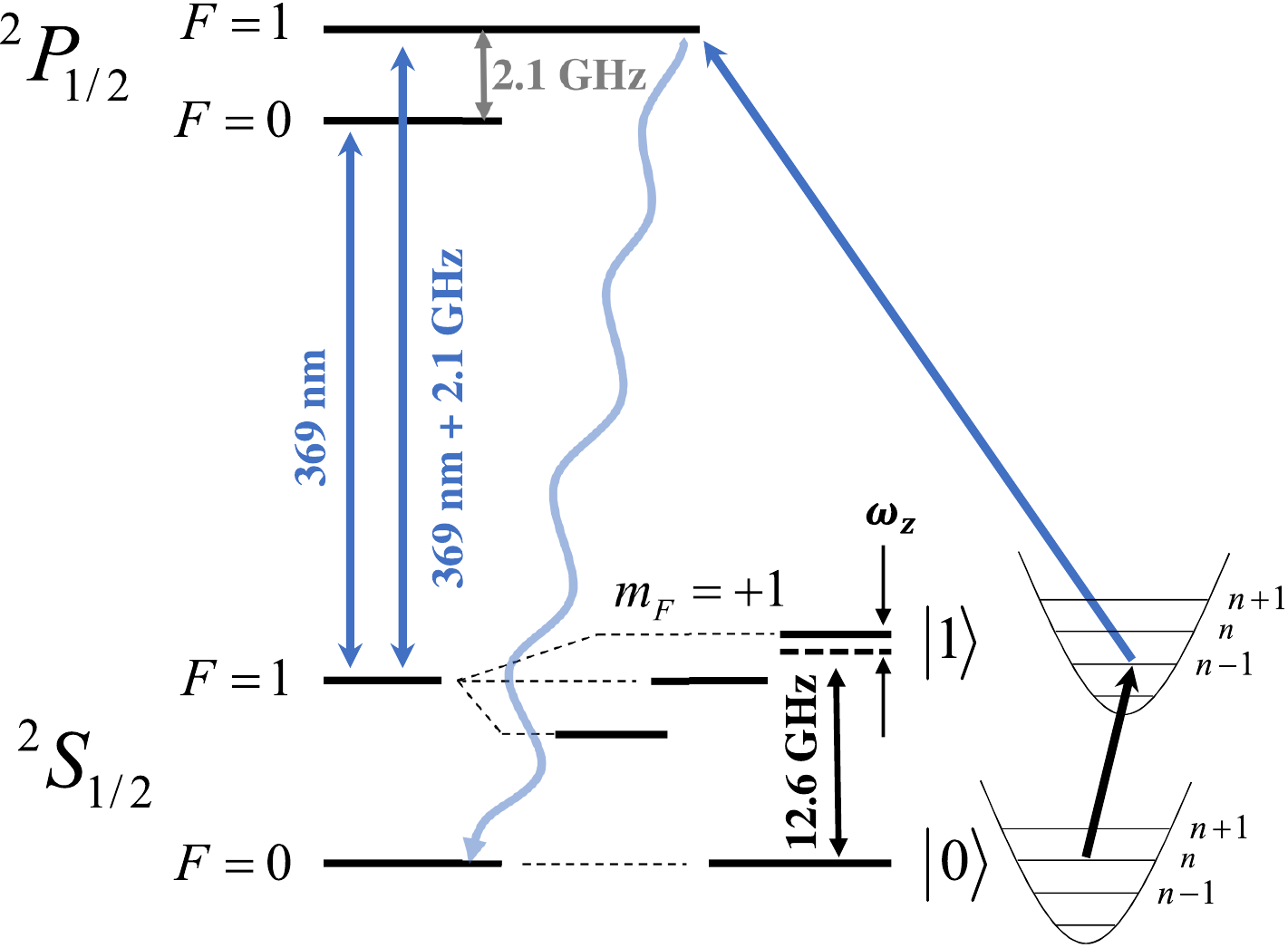}
\caption{(Colour online) Partial energy level diagram of a \ybplus ~ion. First laser light near 369 nm and RF radiation at 12.6 GHz are turned on for Doppler cooling. Then the 369 nm laser is blue detuned by 2.1 GHz and the resonance transition $\ket{0} \leftrightarrow \ket{1}$ is driven by RF radiation that is red detuned by the  frequency of the ions' harmonic motion ($\omega_z$) to perform sideband cooling.}
\label{level-scheme}
\end{figure}

In what follows we describe the principle of RF sideband cooling and discuss the ways to determine the average phonon number. First, a single \ybplus  ion or a Coulomb crystal of two \ybplus ions  is Doppler cooled as described above. Then, to start the sideband cooling cycle, as shown in \fig{level-scheme}, the ion is prepared in state $\ket{0}$ by resonantly driving the \sh (F=1) $\leftrightarrow$ \ph (F=1) transition. The ion is excited to state $\ket{1}$ on the vibrational sideband (first-order in the LDP)using RF radiation red detuned relative to the ionic resonance by the angular vibrational frequency $\omega_z$. Since there is no spontaneous emission between the described qubit levels, the transition \sh (F=1) $\leftrightarrow$ \ph (F=1) is continuously driven to optically repump the ion into the ground state \citep{wunderlich2005}. 

\begin{figure}[]
\centering
\includegraphics[width=0.48\textwidth]{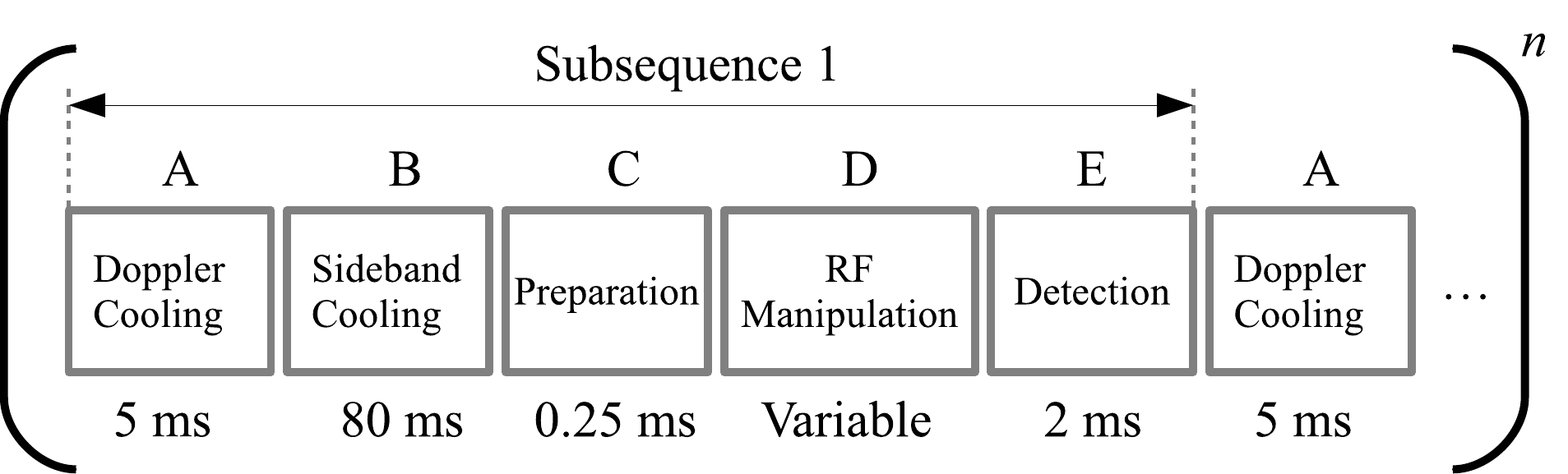}
\caption{(Colour online) The experimental pulse sequence, underlying the measurements to obtain Rabi oscillations and motional sideband spectra after cooling, consists of (A) Doppler cooling, (B) sideband cooling, (C) preparation of qubit state $\ket{0}$, (D) RF manipulation and (E) detection. Subsequence 1 results in a single datum. The next sequences are similar to sequence 1 except that the parameters of the RF manipulation (either the RF pulse duration or the RF frequency detuning) are varied. The complete chain of such sequences is repeated $n$ times, where $n$ is chosen according to the desired level of precision. 
}
\label{pulse-seq}
\end{figure}

A  description of the generic pulse sequence employed for the measurements reported here is shown in Fig.~\ref{pulse-seq}. The experimental pulse sequence consists of several subsequences (labeled subsequence 1 in Fig.~\ref{pulse-seq}), each comprised of Doppler cooling, sideband cooling, preparation of qubit state $\ket{0}$, RF manipulation and detection. The parameters for RF manipulation are varied in each subsequence (see Sec.~\ref{cool-one-ion}) to acquire all data points for diagnosing the motional state of the ion(s). The pulse sequence is then repeated for a sufficient number of times to reduce the statistical error of each data point.

\subsection{Cooling of a single ion}
\label{cool-one-ion}

Two different approaches were followed to extract the average phonon number after sideband cooling. 

\begin{figure}[b]
\centering
	\subfigure[Doppler Cooling]	{
			\includegraphics[width=0.51\textwidth]{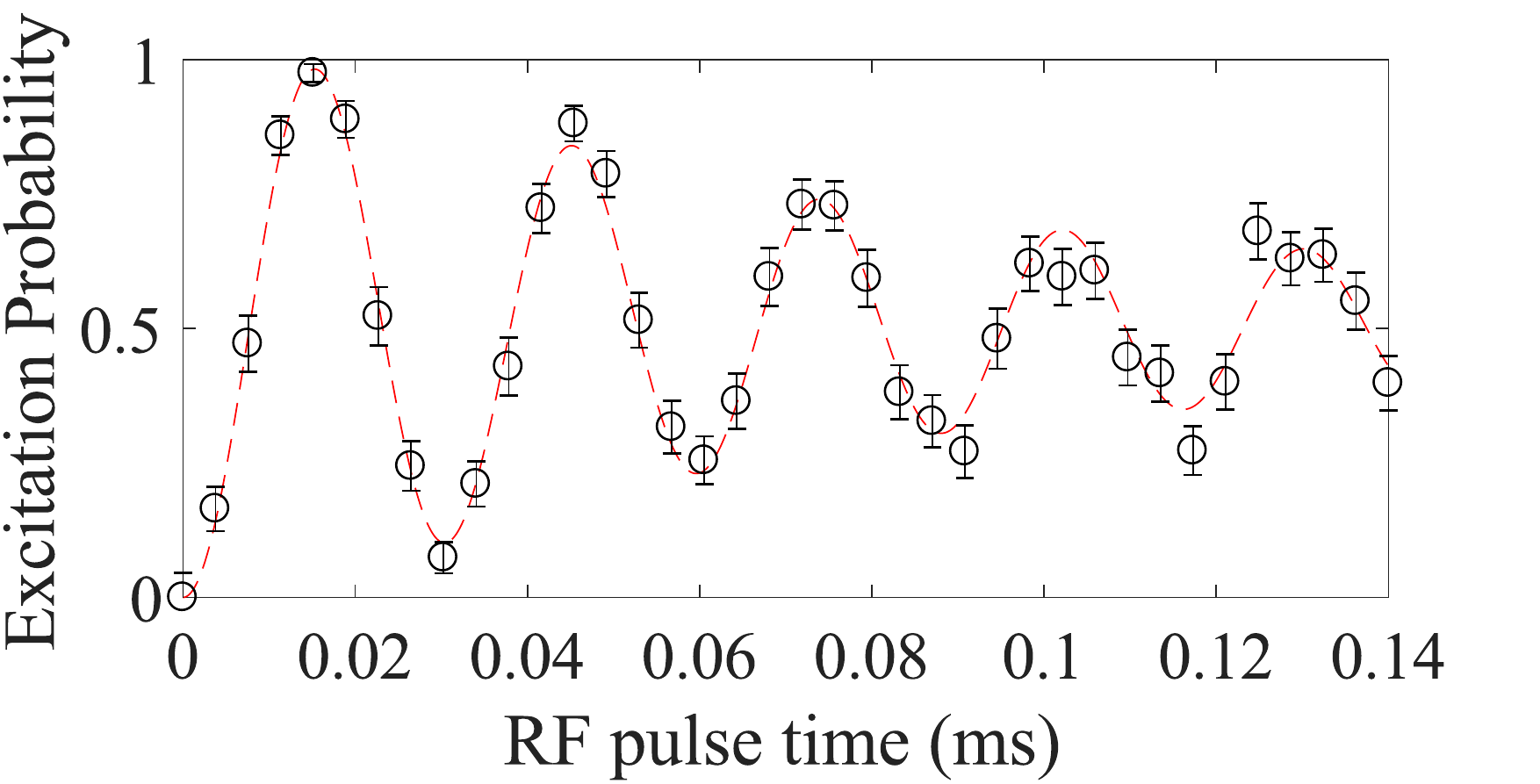}
			\label{rabi-osc-doppler-cooling}
								} 
	\subfigure[Sideband Cooling]{
			\includegraphics[width=0.48\textwidth]{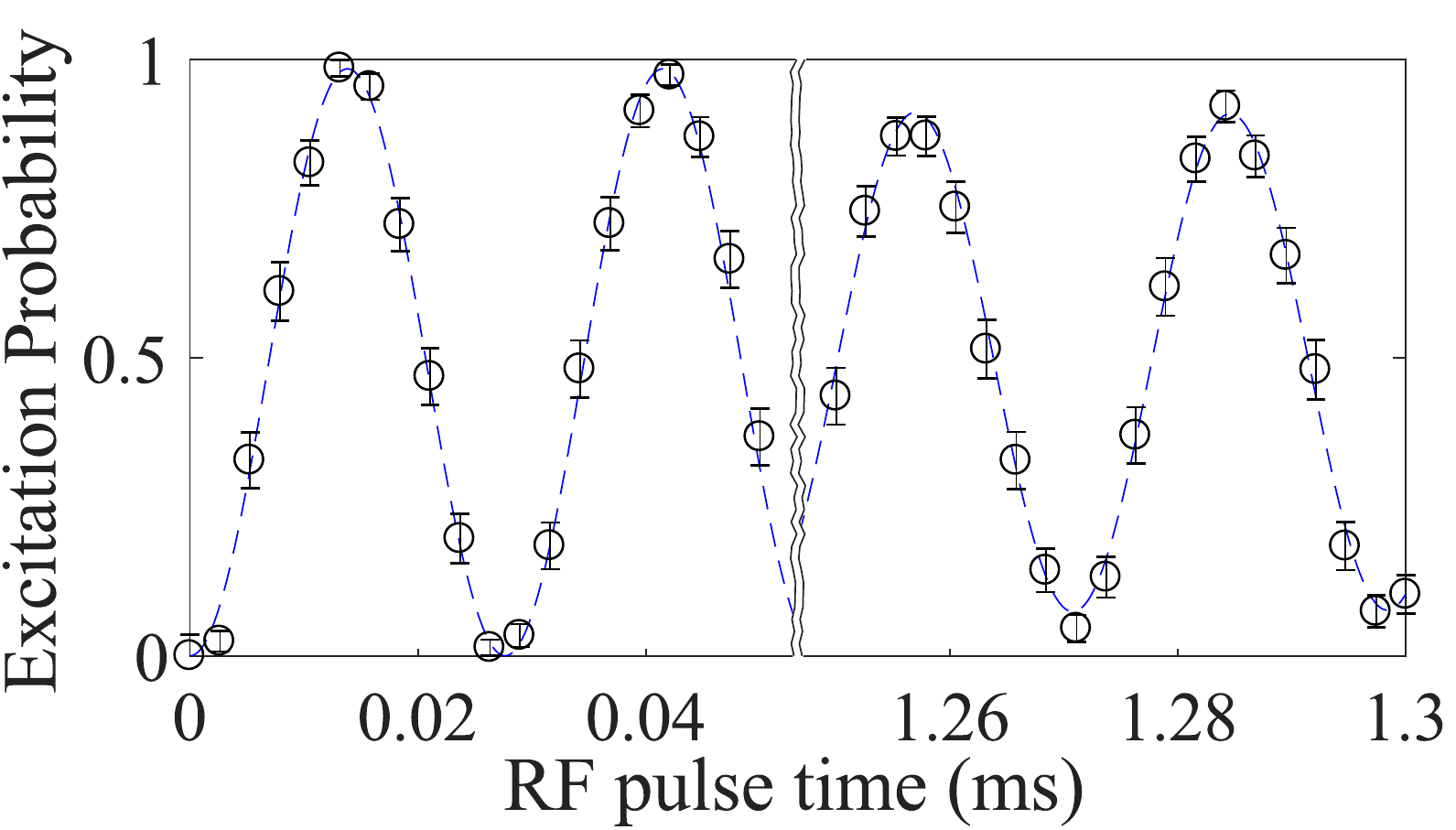}
			\label{rabi-osc-sbc}
								} 
\caption{(Colour online) Determination of the average phonon number by fitting Rabi oscillations on the $\ket{0} \leftrightarrow \ket{1}$ resonance. (a) The ion is only Doppler cooled before the measurement is performed. (b) RF sideband cooling is applied for a duration of 80~ms. Measurements are carried out in two different time windows corresponding to the applied RF pulse duration of 0$-$50 $\mu$s (left) and 1.25$-$1.30 ms (right). The average phonon number deduced from the fits are (a) $\mean{n}=$ 91(5) and (b) $\mean{n}=$ 1.71(15). Each data point represents 125 repetitions. The error bars represent statistical errors within one standard deviation.
}
\label{rabi-osc-doppler-sbc}
\end{figure}

(i) Rabi oscillations are observed by tuning the RF to resonance with the transition $\ket{0} \leftrightarrow \ket{1}$ and by varying the duration of the RF pulses. Since the Rabi frequency depends on the phonon number $n$ (see appendix~\ref{sec:n-decoherence}), damping of Rabi oscillations is observed, if several vibrational energy states are occupied. The average phonon number achieved after Doppler cooling is 91(5), as shown in Fig.~\ref{rabi-osc-doppler-cooling}. 
The laser beam for sideband cooling has an intensity of 0.21(4) W/m$^2$ and is focused to a waist of 144(3) $\mu$m at the location of ion. Sideband cooling is applied for 80 ms which reduces the phonon number to 1.71(15), as shown in Fig.~\ref{rabi-osc-sbc}.

\begin{figure}[]
\centering
		\subfigure{
			\includegraphics[width=0.45\textwidth]{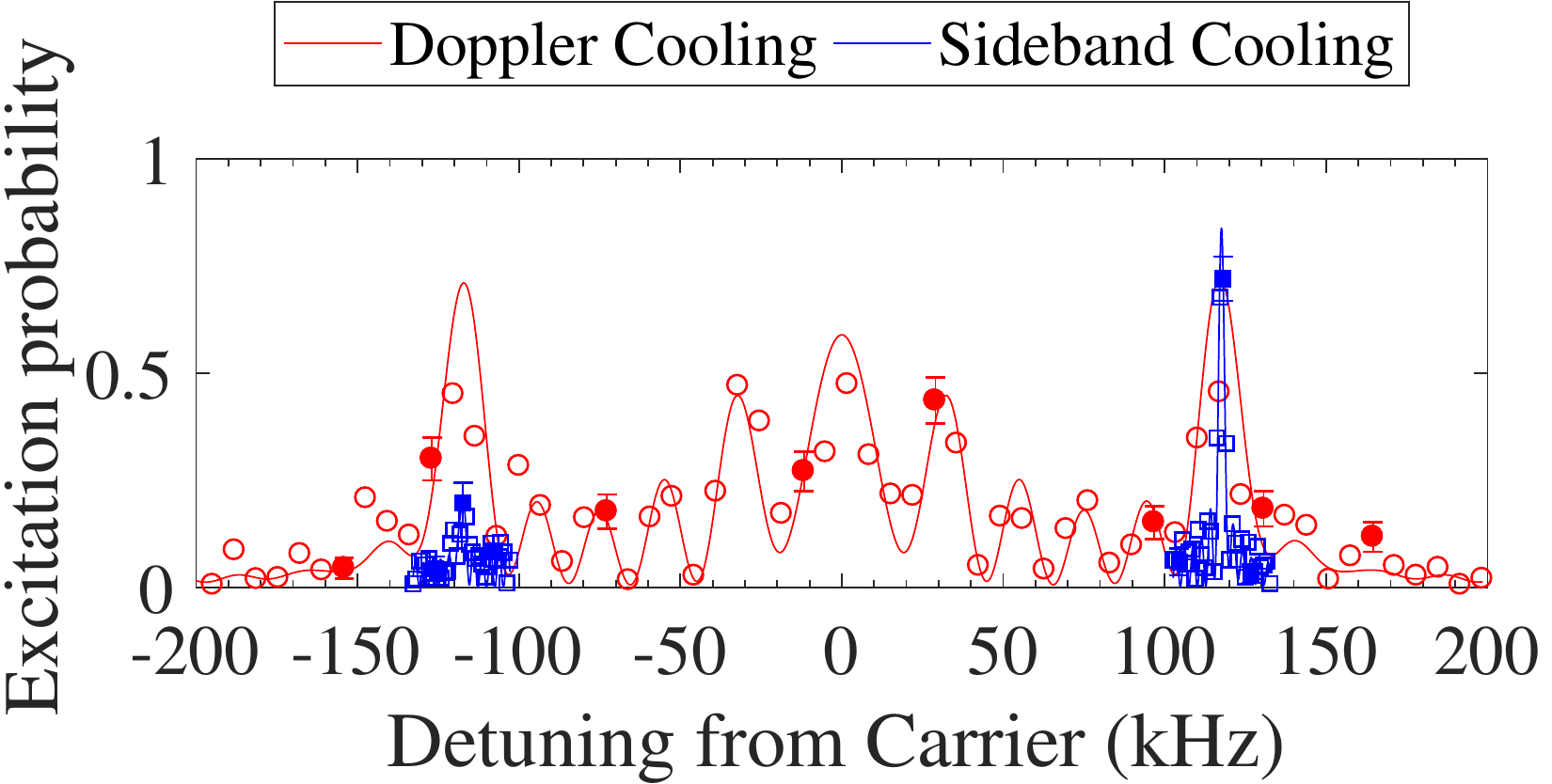}
			\label{fig:heat-cool:compare}
							}	
		\subfigure{
			\includegraphics[width=0.45\textwidth]{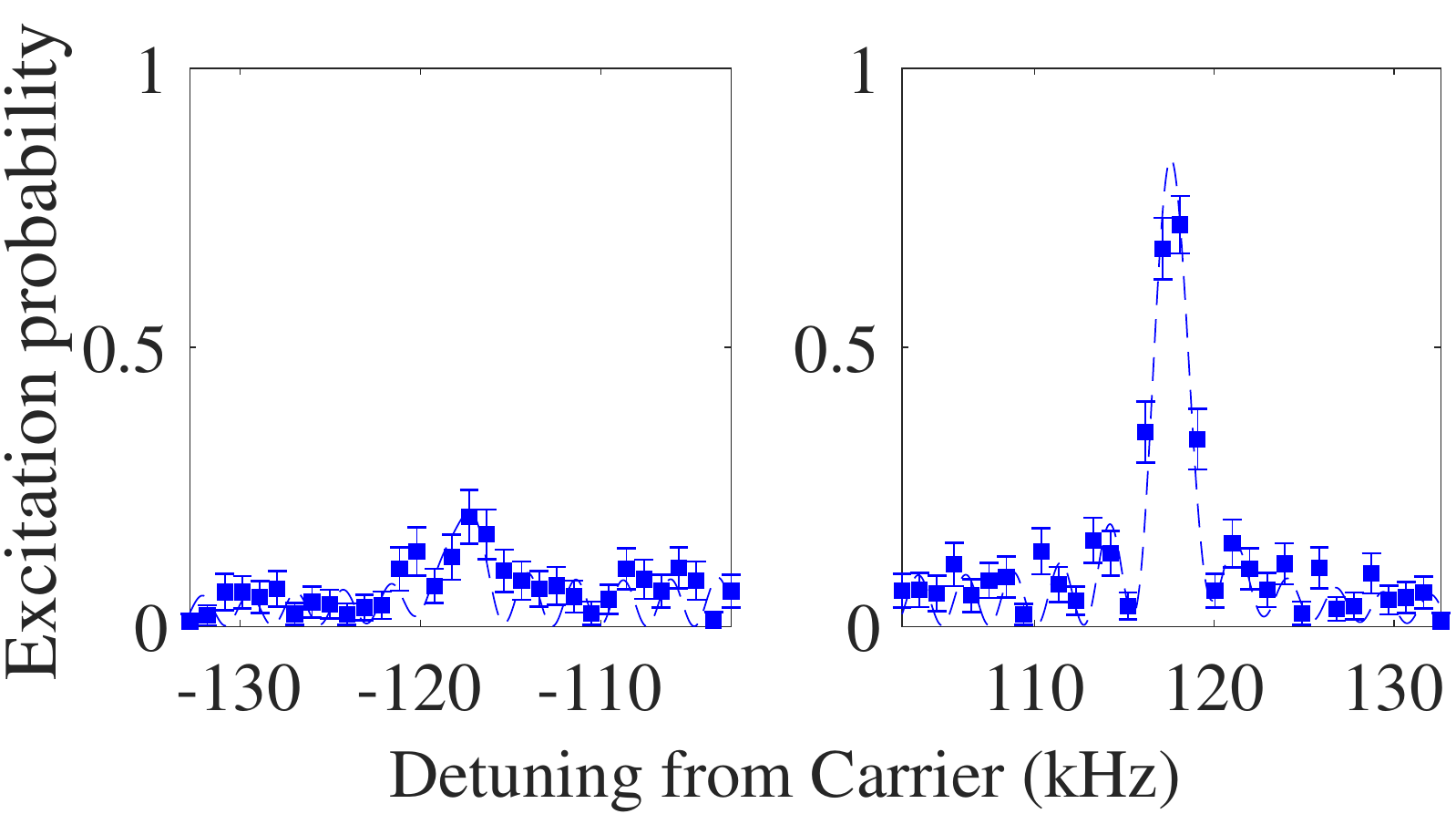}
			\label{fig:heat-cool:cold}
							}	
\caption{(Colour online) A single \ybplus ion's  thermal excitation after Doppler cooling (red data points and red solid line) and after additional RF sideband cooling (red data points and red solid line) determined by measuring and fitting the red and blue motional sidebands of the $\ket{0} \leftrightarrow \ket{1}$ resonance. RF sideband cooling is applied for a duration of 80~ms. The horizontal axis shows the detuning of the RF radiation from the qubit resonance. A RF pulse length of 55~$\mu$s was used for recording the RF-optical double resonance spectrum after Doppler cooling, and a pulse length of  400~$\mu$s was used after additional sideband cooling. Preparation and detection laser light was applied for 250 ~$\mu$s and 2.2 ms, respectively. A comparison of the spectra after Doppler cooling and after additional sideband cooling demonstrates that the red sideband is significantly suppressed after sideband cooling. The average phonon number deduced from the fits are $\mean{n} =$ 65(22) (red) and $\mean{n} =$ 0.30(12) (blue). The sidebands shown in the two lower plots represent a magnified view of the sideband cooling result presented in the upper plot. For clarity, the upper plot shows error bars only for a few selected data points. Each data point represents 100 repetitions. The error bars represent statistical errors within one standard deviation.}
\label{sb-cooling}
\end{figure}

(ii) Instead of varying the duration of the RF pulse, the frequency of the RF pulse is varied and an RF-optical double-resonance spectrum is obtained from measuring the excitation probability as a function of RF frequency.  In order to obtain such a spectrum, 
RF pulses of fixed length and a given frequency are applied close to the $\ket{0} \leftrightarrow \ket{1}$  transition after preparing an ion in state $\ket{0}$. The pulse length is selected such that the probability to excite the red and blue sidebands is enhanced. Then, state selective detection of state $\ket{1}$ takes place. 
Thus, the measured excitation probability as a function of RF frequency yields an RF-optical double resonance spectrum as shown in Fig. \ref{sb-cooling}. The measured vibrational frequency $\omega_z$ is $2\pi \times$ 117.48(11) kHz. The fitting function that allows for extracting the mean vibrational excitation $\mean{n} $ from this spectrum is described in appendix~\ref{sec:n-spectrum}. 

If vibrational states are excited, both the blue and the red sidebands are visible, whereas the resonance of the red sideband is strongly reduced when the ion is cooled  close to the vibrational ground state \citep{wineland1975}. The mean phonon number after Doppler cooling, deduced from this method, is 65(22). Quantifying such a high vibrational excitation using an RF-optical double-resonance spectrum, as opposed to the damping of Rabi oscillations, leads to a larger error, since the phonon number extracted from the spectrum depends on the ratio between the excitation probabilities on the red and blue sidebands. This ratio is close to unity when the vibrational excitation is high and therefore leads to a large error in the inferred phonon number. Reduction of this error would require particularly large number of  repetitions of individual measurements. 

The average phonon number inferred from the RF resonance spectrum after sideband cooling is $\mean{n} =0.30(12)$. This is shown in Fig.~\ref{sb-cooling}
The difference in the deduced phonon numbers after the sideband-cooling cycle, that is for small phonon numbers, following the two approaches described above (\fig{rabi-osc-doppler-sbc} and \fig{sb-cooling}), is attributed mainly to a change in the qubit resonance frequency over time. 
The drift for the magnetic field sensitive qubit is typically 2-3 Hz/s and can accumulate to about 50-100 Hz during the measurement of each data point. A large number of repetitions can therefore result in a reduction of the contrast of Rabi oscillations, and thereby an error is induced in the deduction of the vibrational excitation. On the other hand, even though such drifts also cause a reduction in the excitation probabilities of the motional sidebands, the quantification of vibrational excitation remains largely unaffected as the latter depends on the ratio of excitation probabilities of the red and blue sidebands.
Therefore, the phonon number deduced by observing the decay of Rabi oscillations is viewed as an upper limit for the actual mean phonon number.

\begin{figure}[]
	\centering
	\includegraphics[width=0.49\textwidth]{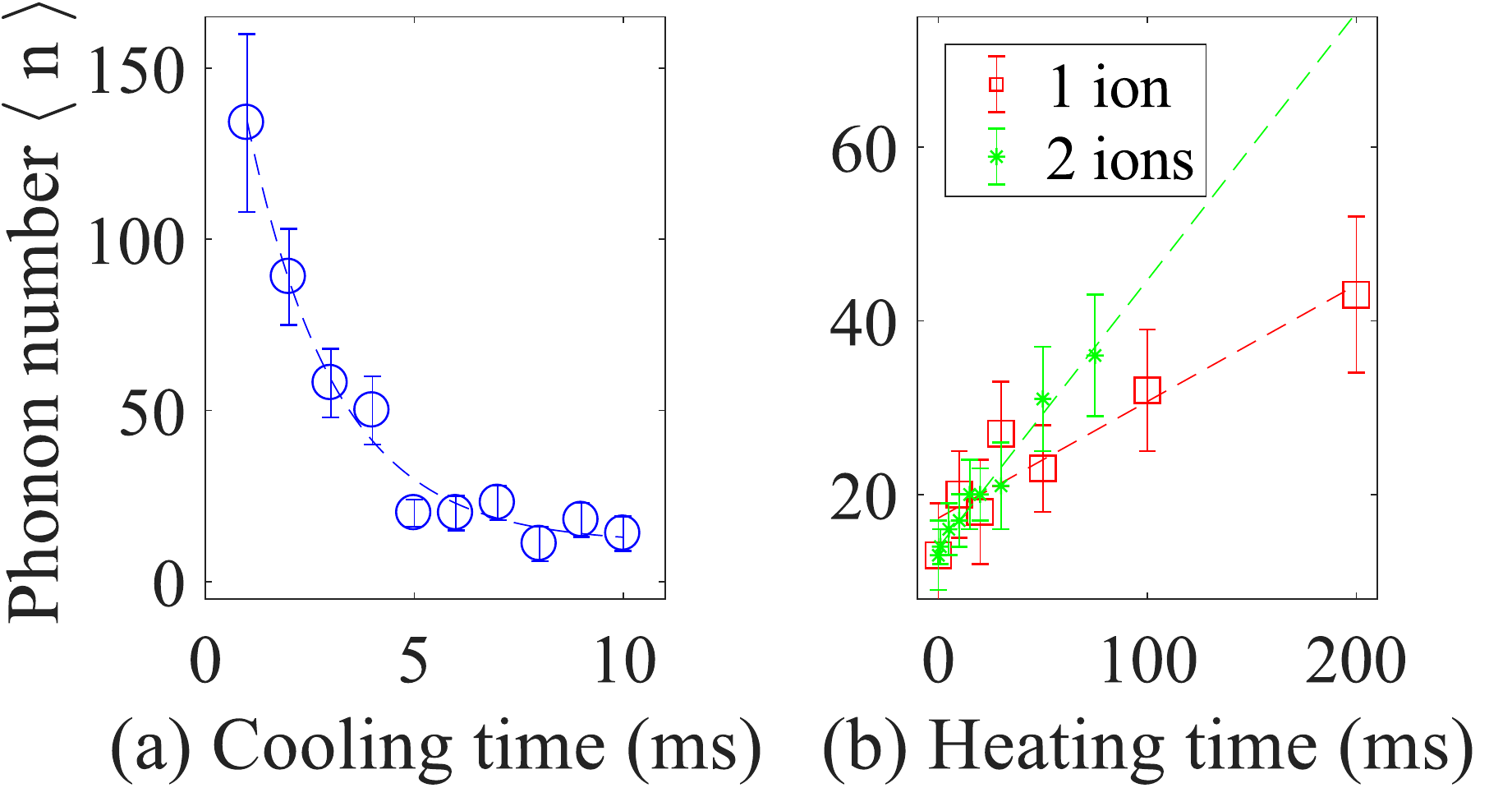}
\caption{(Colour online) Determination of the cooling (a) and heating (b) rates for a single ion. (a) The average phonon number is plotted as a function of the sideband cooling time after the ion was initially Doppler cooled. From the exponential fit, the deduced sideband cooling rate is $0.47(6)$~ms$^{-1}$. (b) After $15$~ms of sideband cooling, the cooling laser and RF were turned off and the ion was left in the trap for a given time without cooling. During this time the temperature of the ion increases again. From the linear fit, the deduced heating rate is $0.13(2)~$ms$^{-1}$ for a single ion system (square). For a two ion system (star), the sideband cooling is applied for 50 ms on the COM mode of the refrigerant ion. The deduced heating rate from the corresponding linear fit is 0.30(2)~ms$^{-1}$. Each data point represents 35 repetitions. The error bars represent statistical errors within one standard deviation.}
\label{fig:heat-cool}
\end{figure}

The theoretical prediction for the lowest attainable mean phonon number with the help of sideband cooling  can be determined by $\mean{n} = {R_H} / (R_{SBC}-R_H)$, with the heating rate, $R_H$ of the ion, and the sideband cooling rate,  $R_{SBC}$,  respectively \citep{stenholm1986}. We measure the mean vibrational excitation $\mean{n}$ as a function of the sideband cooling time by observing Rabi oscillations as is outlined above. The result shown in Fig.~\ref{fig:heat-cool}~(a) indicates an exponential decay of the phonon number with time and a sideband cooling rate of $ R_{SBC} = 0.47(6)$~ms$^{-1}$ is deduced from an exponential fit of the experimental data. The heating rate is measured by first  sideband cooling an ion for 15 ms and then let the ion heat up for varying durations before observing Rabi oscillations. The result shown in \fig{fig:heat-cool}~(b) indicates a linear rise in phonon number and the heating rate of $ R_{H} = 0.13(2)$~ms$^{-1}$ is determined from this result. From these two rates, the theoretical limit of sideband cooling is estimated to be $\mean{n} = 0.37 (10)$. Our experimental result, showing a mean phonon number $\mean{n} = 0.30(12)$ after sideband cooling (\fig{sb-cooling}), agrees well with this theoretical limit. 

Fig.~\ref{fig:heat-cool}~(a) shows a higher phonon number after Doppler cooling than in the results quoted above. This was caused by a higher heating rate at the time the data shown in Fig.~\ref{fig:heat-cool}~(a) was taken. The heating rate was subsequently reduced by installing additional low-pass filters to suppress noise components in the DC voltages applied to the trap electrodes.

\begin{figure}[]
	\centering	
	\includegraphics[width=0.48\textwidth]{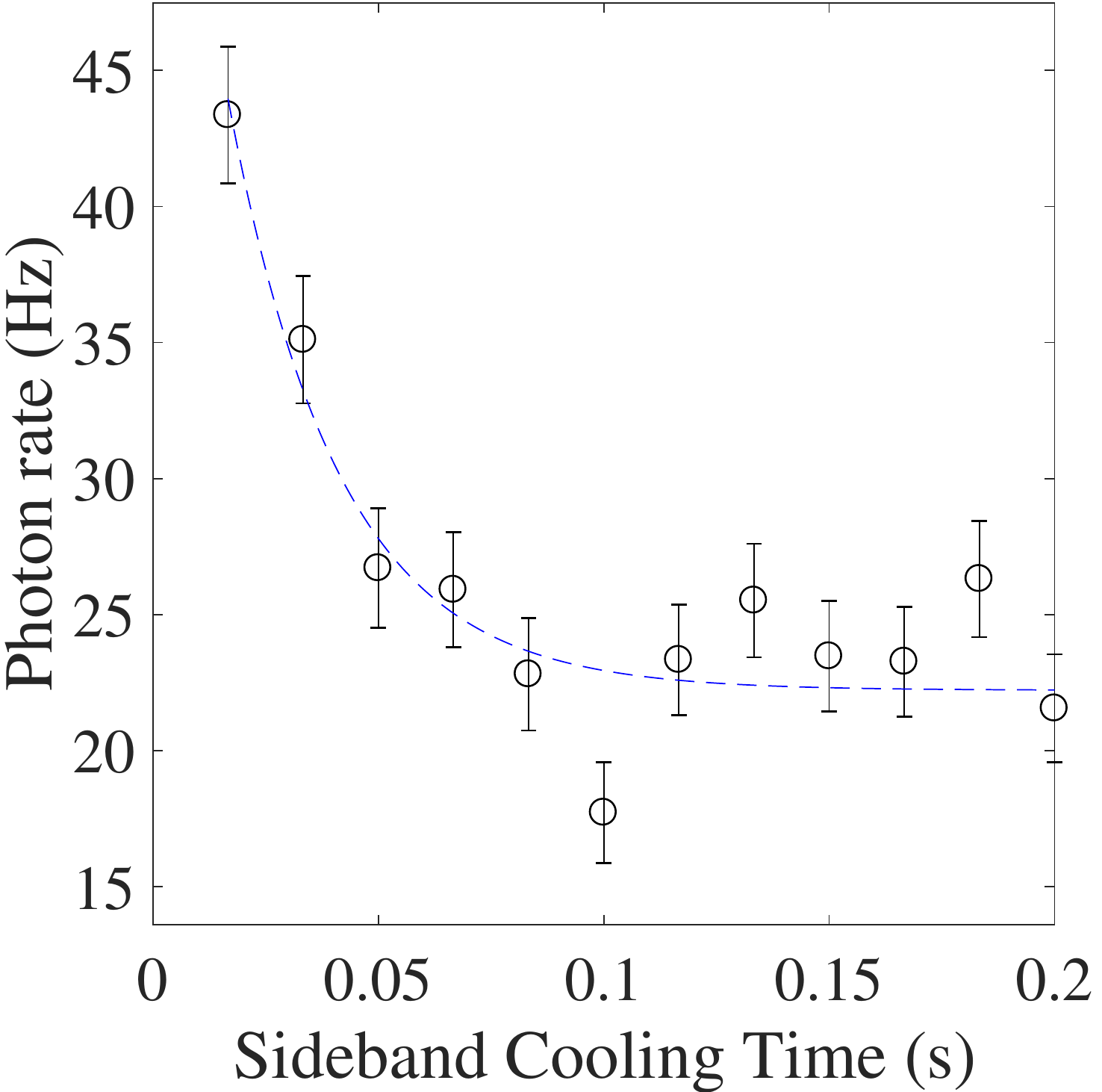}
	\caption{ (Colour online) Determination of optimal sideband cooling duration by observing the photon count rate during the sideband cooling process. The RF frequency is set at the red sideband transition. The decay constant deduced from the fit is $25(8)^{-1}$~ms$^{-1}$. Each data point represents 1200 repetitions. The error bars represent statistical errors within one standard deviation.}
	\label{fig:sbc-process-rate}
\end{figure}

The optimal duration of sideband cooling was obtained by measuring the fluorescence photon count rate during the sideband cooling process. These measurements were repeated for different durations of sideband cooling and an exponential decay of photon count rate is observed (Fig.~\ref{fig:sbc-process-rate}). As the mean phonon number decreases in time, the ion is excited less often by the red-detuned RF radiation, leading to a reduced photon count rate. 
The optimal duration of sideband cooling is chosen as 80~ms where the exponential decay closely approaches its asymptote.

\subsection{Cooling of two ions}
\label{cool-two-ions}

The sideband cooling method as discussed above for a single ion is extended to a two \ybplus ion Coulomb crystal. A comprehensive theoretical description for laser cooling of two ions can be found in Ref. \citep{morigi1999}. Here, we demonstrate sympathetic cooling of a Doppler cooled target ion that is part of a two-\ybplus ion crystal by sideband cooling  a second ion,  the refrigerant ion. Previous demonstrations of sympathetic cooling employing laser radiation made use of two identical  ions (e.g., \citep{rohde2001}), or of two different isotopes of the same ion species (e.g.,\citep{ballance2015}, or of two different ion species (e.g., \citep{barrett2003, guggemos2013}). In this work, we demonstrate  complementary results using RF sideband cooling.

In the presence of a magnetic field gradient the two \ybplus ions are distinguished by their resonance frequency on the $\ket{0}\leftrightarrow\ket{1}$ transition with a frequency difference of 3.7 MHz. The two-ion crystal has two normal modes of vibration: a centre-of-mass (COM) mode where the ions oscillate in phase, and a stretch (STR) mode where the ions oscillate 180$^\circ$ out of phase.
The effective LDP for two ions in our experiment is ${\eta_{eff}}/{\sqrt{2}}$ = 0.0254 (COM) and ${\eta_{eff}}/{\sqrt{2\sqrt{3}}} = 0.0193$ (STR).
In order to efficiently cool the vibrational modes the frequencies of both modes are  determined to be $\omega_{\mathrm{COM}}$ = $2\pi \times$ 117.23(5) and $\omega_{\mathrm{STR}}$ = $2\pi \times$ 209.54(14) kHz, 
following the same procedure as described in Sec.~\ref{cool-one-ion}.
With $\omega_z$ = $2\pi \times$ 117.48(11) kHz as the axial secular frequency of a single ion, the vibrational mode frequencies are expected to be $\omega_{\mathrm{COM}}$ = $\omega_z$ = $2\pi \times$ 117.48 kHz and $\omega_{\mathrm{STR}}$ = $\sqrt{3}\omega_z$ = $2\pi \times$ 203.48 kHz. The observed deviation is possibly caused by an imperfect compensation of micromotion. Any residual micromotion can displace the ions from the trap centre resulting in a coupling between the axial and radial modes \citep{roth2007}. This effect leads to a shift of the axial vibrational mode frequencies \citep{goeders2013, barrett2003}. In the following we report RF sideband cooling of both axial vibrational modes, applied and measured independently for the COM mode and the STR mode. 

Immediately after initial Doppler cooling, where the mean frequency between the two RF resonance frequencies for two ions on the $\ket{0} \leftrightarrow \ket{0'}$ transition is used, the mean phonon numbers for the COM and the STR modes are measured by recording RF-optical double resonance spectra for the respective modes. $\mean{n_\mathrm{COM}}$ = 64(23) and $\mean{n_\mathrm{STR}}$ = 11(5) are inferred from the red and blue sidebands for the respective frequency modes (see Table~\ref{tab:cooling2ions}). The significant difference in the thermal excitation for the COM mode compared to that of the STR mode is likely due to a larger coupling of the former mode to stray electric fields \citep{marinescu2011, home2013} resulting in a higher heating rate for the COM mode, and thus $\mean{n_\mathrm{COM}} > \mean{n_\mathrm{STR}}$. 

\begin{figure}[]
\centering
\includegraphics[width=0.5\textwidth]{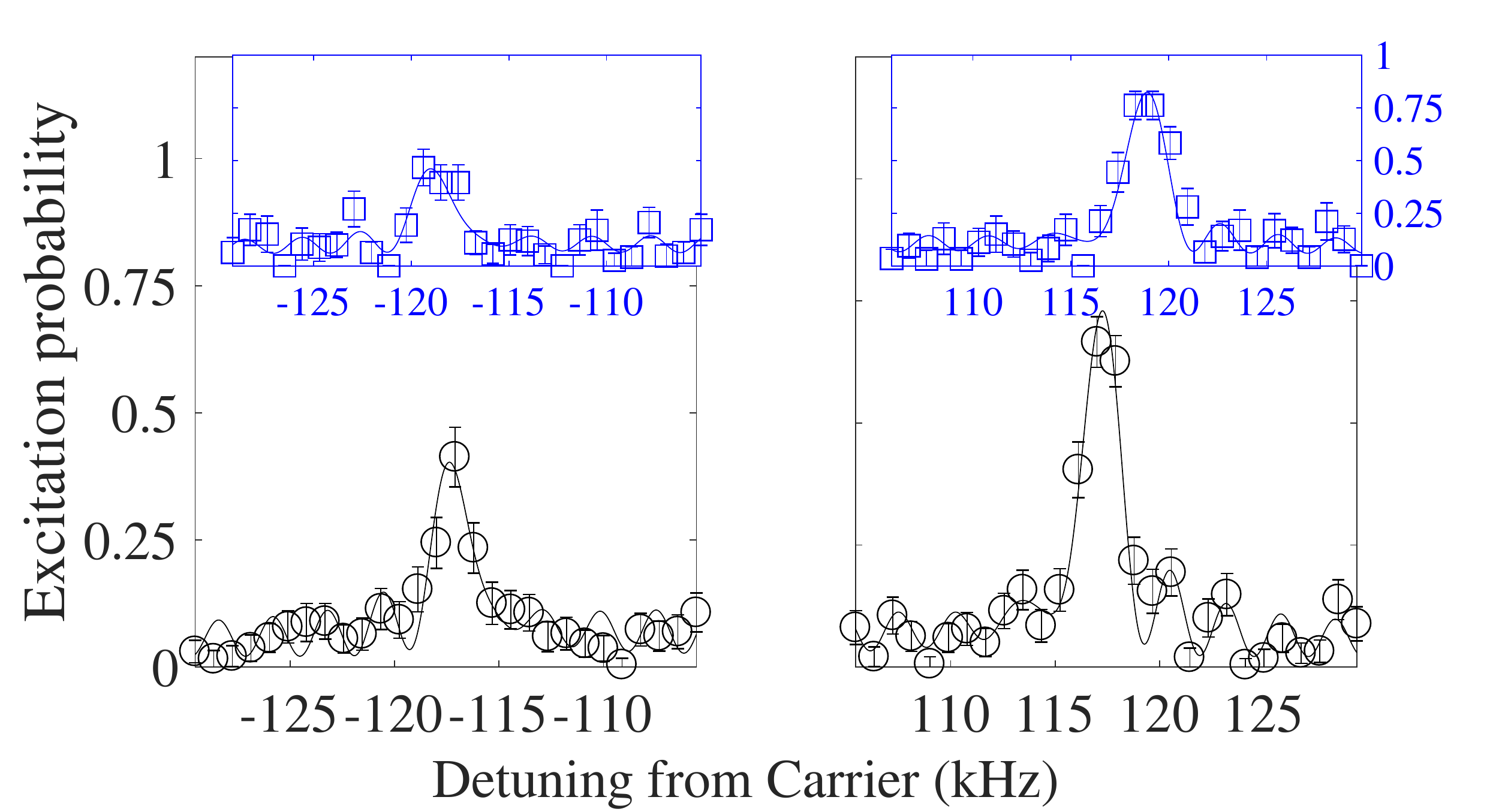}
\caption{(Colour online) Demonstration of sympathetic cooling using two identical isotopes of the same ion species. The red sidebands (left peaks) and blue sidebands (right peaks) of the refrigerant ion (circles) and the target ion (squares) are shown. The average phonon numbers inferred from the fits are $\mean{n} =$ 1.1(4) (refrigerant) and $\mean{n} =$ 1.0(6) (target). Each data point represents 100 repetitions. The error bars represent statistical errors within one standard deviation.}
\label{sympathetic-cooling}
\end{figure}

We apply RF sideband cooling only to one ion, the refrigerant ion, by tuning the RF radiation to its red sideband resonance. By cooling the COM mode for 80 ms (with an intensity of the repump laser of 0.57(4) W/m$^2$), a mean phonon number of $\mean{n_\mathrm{COM}}$ = 1.1(4) can be inferred from the sidebands of the refrigerant ion (circular markers in \fig{sympathetic-cooling}). When sideband cooling is applied to the  STR mode (repumper intensity of 8.16(4) W/m$^2$) a  mean phonon number $\mean{n_\mathrm{STR}}$ = 3.2(1.1) is achieved (see also Table~\ref{tab:cooling2ions}). These results are obtained by cooling and diagnosing the same ion.

Instead of probing the refrigerant ion, the target ion is now probed and a mean phonon number of $\mean{n_\mathrm{COM}}$ = 1.0(6) is inferred from the corresponding sidebands (square markers in \fig{sympathetic-cooling}). Similarly, by cooling only the STR mode for 80 ms  a mean  phonon number $\mean{n_\mathrm{STR}}$ =  4.2(1.7) is inferred from the sidebands of the  target ion (see Table~\ref{tab:cooling2ions}). The phonon numbers observed from different ions are  compatible within the error bars demonstrating efficient sympathetic cooling.

\begin{table}[]
\caption{Summary of average phonon number achieved after Doppler and sideband cooling (SBC) applied on a two ion system. A blank entry indicates that no experimental result is available.}
{
\begin{tabular}{ l c c c c }
\hline
Cooling (on ion) & \multicolumn{2}{c}{Refrigerant ion} & \multicolumn{2}{c}{Target ion} \\  \cline{2-5}
 				 						     & COM & STR & COM & STR \\ 
\hline
Doppler cooling (both) 						 & 64(23) & 11(5) & 62(20) & 15(8) \\ 
SBC on COM (Refrigerant) 							 & 1.1(4) & - & 1.0(6) & -  \\ 
SBC on STR (Refrigerant) 						     & -  & 3.2(1.1) & -  & 4.2(1.7)  \\ 
\hline
\end{tabular} 
}
\label{tab:cooling2ions}
\end{table}

Additionally, the heating rate is measured for a two ion system, initially after sideband cooling for 50 ms on the COM mode of the left ion and then allowing the ions to heat up by pausing the sideband cooling for a given amount of time. Finally the RF-optical double resonance spectra for different heating times are recorded. The deduced heating rate with respect to the COM mode of the refrigerant ion is 0.30(2) ms$^{-1}$ as shown in Fig.~\ref{fig:heat-cool}~(b).


\section{Conclusion}
\label{sec:conclusion}

To summarise, we have demonstrated near-ground state cooling of a single trapped atomic ion exposed to a static magnetic field gradient, employing a simple RF sideband cooling technique. The sideband cooling method is demonstrated at a low secular trap frequency as is also often encountered in neutral atom traps. Given the widespread use of hyperfine transitions in many neutral atom and ion experiments, and the easy availability of  off-the-shelf commercial RF tools, combined with its cost effectiveness when compared to Raman laser set-ups, the method presented here is expected to be useful for future experiments in the ion and atom trapping communities. Furthermore, the first demonstration of sympathetic sideband cooling in a two ion crystal using RF radiation complements the conventional approaches of laser cooling. The lowest attainable phonon number in proof-of-principle experiments reported here can be reduced further by lowering the heating rate.

\section*{Acknowledgments}
The authors acknowledge funding from Deutsche Forschungsgemeinschaft and from Bundesministerium f{\"u}r Bildung und Forschung (FK 16KIS0128). G.~S.~G. also acknowledges support from the European Commission's Horizon 2020 research and innovation programme under the Marie Sk\l{}odowska-Curie grant agreement number 657261.

\bibliographystyle{apsrev4-1}
\bibliography{sbc-references}


\section*{Appendices}
\label{sec:appendices}
\appendix

\section*{Determine $\langle n\rangle$ by damping of Rabi oscillation for a single ion}
\label{sec:n-decoherence}

To estimate the average phonon number $\langle n\rangle$ from damping of Rabi oscillation for a single ion we need to determine the probability $P_{\ket{1}}$ to find the ion in state $\ket{1}$ which is given by \cite{walls1994}
\begin{equation}
P_{\ket{1}}=\frac{1}{2}\left(1-\sum_{n=0}^N p_n(T),
\cos{(\Omega_{n,n}t)}\right)\label{eq:P_res}
\end{equation}
with the pulse duration $t$ and the thermal distribution $p_n(T)$
\begin{equation}
p_n(T)=\frac{1}{\langle n\rangle_T +1}\left(\frac{\langle n\rangle_T}{\langle n\rangle_T +1}\right)^n, 
\end{equation} 
of the phonon number at temperature $T$. In the non-linear coupling regime, the Rabi frequencies $\Omega_{n,n+k}$ of the transition $\ket{0,n}\leftrightarrow \ket{1,n+k}$ are determined by \cite{vogel1995}

\begin{equation}
\Omega_{n,n+k}=\E^{-\eta^2/2}\eta^k\Omega_0 L_n^k(\eta^2)\sqrt{\frac{n!}{(n+k)!}},
\label{eq:omega_n_n+k}
\end{equation} 
with the effective Lamb-Dicke parameter (LDP) $\eta$ \cite{mintert2001} and the Laguerre polynomial 
\begin{equation}
L_n^k(x)=\sum_{j=0}^n(-1)^j\binom{n+k}{n-j}\frac{x^j}{j!},
\end{equation}
of order $n$. The effective LDP caused by the magnetic gradient can be  determine by 
\begin{equation}
\eta_{eff}=\frac{\zeta \Delta z g_F \mu_B \partial_z |B|}{\hbar
\omega_z},
\end{equation} 
with the expansion coefficient of the displacement of the ion $\zeta =
1$, the spatial expansion of the wave function
$\Delta z=\sqrt{\frac{\hbar}{2m\omega_z}}$, the Land\'{e} g-factor $g_F
= 1$, the magnetic field gradient $\partial_z |B|=19$ T/m,
Bohr magneton $\mu_B$, the ion-mass $m = 2.84 \times
10^{-25}\:$kg and the axial trap frequency
$\omega_z=2\pi \times 117.48\:$kHz.  The resulting LDP in our experiment is
given by $\eta_{eff} = 0.036$.

The average phonon number is determined by measuring the probability $P_{\ket{1}}$ and fitting it with the theoretical predicted $P_{\ket{1}}$.

\section*{Determine $\langle n \rangle$ by resonance spectrum}
\label{sec:n-spectrum}

The second way to determine the phonon number is to leave
the duration of the RF excitation constant and vary the
frequency of the excitation instead. After the ion was prepared in
the ground state $\ket{0}$ it has a probability $P_{\ket{1}}$ to
end up in the excited state $\ket{1}$ via excitation
of the carrier and the sidebands. In the case of $\Omega_0 \ll
\omega_\mathrm{mode}$, $P_{\ket{1}}$ is the sum over all probabilities to be
excited via the carrier-transition ($P_C$), the red ($P_R^k$) and blue
sidebands ($P_B^k$). For the carrier, the probability
to find the ion in the excited state is given by
\begin{equation}
P_C(\delta,T)=\sum_n^\infty
p_n(T)\frac{\Omega_{n,n}^2}{\Omega_{n,n}^2+\delta^2}\sin^2{\left(\frac{\sqrt{\Omega_{n,n}^2+\delta^2}}{2}t\right)},
\end{equation}
where $\delta$ is the detuning of the RF from the resonance in units of the angular frequency. For the blue sideband of $k^{\mbox{th}}$
order, which has a detuning of $\delta - k\omega_\mathrm{mode}$ the
excitation probability is given by
\begin{widetext}
\begin{equation}
P_B^k(\delta,T)=\sum_n^\infty
p_n(T)\frac{\Omega_{n+k,n}^2}{\Omega_{n+k,n}^2+(\delta-k\omega_\mathrm{mode})^2}\sin^2{\left(\frac{\sqrt{\Omega_{n+k,n}^2+(\delta-k\omega_\mathrm{mode})^2}}{2}t\right)},
\end{equation}
\end{widetext}
where $\Omega_{n+k,n}$ is adapted from \eq{eq:omega_n_n+k} with $\eta_{eff}$ of $\mathrm{COM} = $ 0.025 and $\mathrm{STR} =$ 0.019.

The probability of excitation through the red sideband is proportional to the excitation through  the blue sideband. However, it is decreased by the factor $(\langle n\rangle /(\langle n\rangle  +1))^k$. As a consequence, it is determined by
\begin{equation}
P_R^k(\delta,T)=\left(\frac{\left<n\right>_T}{1+\left<n\right>_T}\right)^k
P_B^k(-\delta,T).
\end{equation}
To determine the average phonon number we fit the experimentally measured 
spectrum with the function
\begin{equation}
P_{\ket{1}}(\delta,T)=P_C(\delta,T)+\sum_{\mathrm{mode}}\sum_{k=1}^2\left(P_R^k(\delta,T)+P_B^k(\delta,T)\right).
\end{equation}
Here, we considered only the first two orders,
since higher orders cannot be distinguished from noise in our
experiment. Furthermore, for low phonon numbers, the calculation with the first few orders is adequate. 


\end{document}